\documentclass{article}
\usepackage{e,amsmath}
\usepackage{graphics,color,floatflt,epsfig}
\usepackage[letterpaper,colorlinks,breaklinks,bookmarks]{hyperref}
\begin{document}
\renewcommand{\Re}{\hbox{Re\,}}
\renewcommand{\Im}{\hbox{Im\,}}
\branch{D}
\DOI{123}                       
\idline{A}{x, x--x}{1}         
\editorial{}{}{}{}              
\title{Four-wave mixing at maximum coherence
and eliminated Doppler broadening controlled with the driving fields}
\author{A. K. Popov\inst{1,2}\and Alexander  S. Bayev\inst{1}\and Thomas F.
George\inst{3}\and\\ and  Vladimir M. Shalaev\inst{1,4} }
\institute{Institute for Physics, Russian Academy of Sciences,
Krasnoyarsk, 660036, Russia\\
e-mail: \href{mailto:lco@iph.krasn.ru}{lco@iph.krasn.ru}
\and Krasnoyarsk State University and Krasnoyarsk State Technical University\\
e-mail: \href{mailto:popov@ksc.krasn.ru}{popov@ksc.krasn.ru} \and Office of the
Chancellor / Departments of Chemistry and Physics \& Astronomy, University of
Wisconsin-Stevens Point, Stevens Point, WI 54481-3897, USA\\
email: \href{mailto:tgeorge@uwsp.edu}{tgeorge@uwsp.edu} \and Department of Physics,
New Mexico
State University, Las Cruces, NM 88003-8001, USA\\
e-mail: \href{mailto:vshalaev@nmsu.edu}{vshalaev@nmsu.edu}}
\titlerunning
\authorrunning
\maketitle
\sloppy
\begin{abstract}

New feasibity of coherent quantum control of four-wave mixing processes in a resonant
Doppler-broadened medium are studied. We propose a technique which enables one to
enhance the quantum efficiency of nonlinear optical conversion. At the same time, it
allows one to decrease the required intensities of the fundamental beams compared to
those necessary in the approach based on coherent population trapping. The major
outcomes of the analysis are illustrated with numerical simulation addressed within a
practical medium.
\end{abstract}
\section{Introduction}

The concept of quantum coherence and interference plays an important role in resonant
nonlinear optics and spectroscopy \cite{Rau}. Coherent coupling of laser radiation
with multilevel quantum schemes has been used to manipulate energy level populations,
nonlinear-optical response, refraction and absorption of a resonance medium
\cite{QC}. Much attention has been shown to four-wave mixing (FWM) processes with
maximal coherence on a Raman transition \cite{NOMC} based on the effect of coherence
population trapping (CPT) \cite{CPT}, which allows a dramatic increase of conversion
efficiency. In far-from-degenerate schemes, substantial Doppler broadening of the
resonant quantum transitions gives rise to inhomogeneous coupling of the driving
fields with the atoms in different velocity intervals. Therefore, under relatively
low intensities, only a small fraction of the atoms can be concurrently coupled with
both the fields driving the Raman transition and, as a consequence, the CPT
conditions can not be fulfilled for the entire thermal velocity interval. This
fundamental limitation in resonant nonlinear optics of gases substantially reduces
the efficiency of the corresponding processes. The common methods of Doppler-free
(DF) nonlinear spectroscopy entail ladder energy level schemes and equal frequency
counter-propagating beams. This usually leads to large detunings from the
intermediate resonance and consequently a decrease of the nonlinear susceptibility.
Besides that, the method can not be applied to FWM processes in ladder schemes, due
to the phase-matching requirements, and in Raman-type schemes due to the difference
in frequencies of the coupled fields.

The technique, allowing one to overcome this obstacle by
compensating for Doppler frequency shifts with light-induced ac-Stark
shifts and thus enabling one to couple a wide velocity interval
concurrently, has been proposed in \cite{FEOK}. This was developed in
further publications \cite{TAL} and most recently in publications
\cite{BAY} addressing FWM under conditions where CPT is not
possible. Related effects in the absorption index were also
investigated recently \cite {ABS}.

We further mean CPT as the process where two driving fields form nearly equal
velocity-averaged populations of the levels at Raman transitions, while the
velocity-averaged population of the intermediate state is much less. Along with this
absorption and amplification indices for the couple fields reach minimum. On the
other hand coherence in the Raman transition, which is origin of FWM, approaches
maximum.  That allows one to accomplish a maximum of the conversion efficiency. The
conditions to achieve CPT may differ from those required for the DF Raman-type
resonance. \textcolor{red}{This present paper is aimed at investigating what are the
most favorable conditions to accomplish effective FWM frequency conversion of
radiation at a low fundamental radiation level.} This also addresses the problem
``nonlinear optics at the level a few photons per atom'' \cite{HAR}.

\section{Theoretical}
\subsection{Atomic coherence and energy level populations}
\begin{floatingfigure}{0.5\textwidth}
\begin{center}
\includegraphics[width=.45\textwidth]{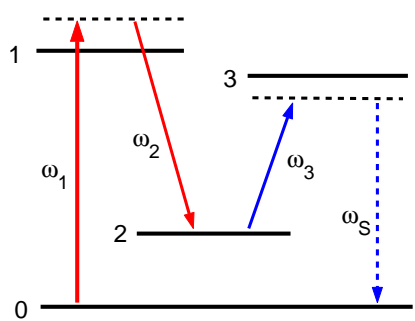}
\end{center}
\vspace{-10pt} \caption{\label{fig1}Energy level configuration}
\end{floatingfigure}
Consider a Raman-type scheme of FWM,  $\omega_S = \omega_1 - \omega_2 + \omega_3$,
depicted in Fig. \ref{fig1}, where $E_1$ and $E_2$ with frequencies $\omega_1$ and
$\omega_2$ are traveling waves of the driving fields.  The field $E_3$ with the
frequency $\omega_3$ and generated radiation $E_S$ with the frequency $\omega_S$ are
assumed weak (i.e., not changing the level populations). All waves are co-propagated.
Initially, only the lower level $\it 0$ is populated. The steady-state solution for
the off-diagonal elements of the density matrix (atomic coherence) can be found in
the same form of traveling waves as the the resonant driving radiations. Then the
equations for the density matrix amplitudes to the lowest order in $E_{3,S}$ can be
written as
\begin{eqnarray}
& P_{01}r_{01}=-i\left[G_{01}(r_{1}-r_{0})- G_{21}r_{02}\right],\,
 P_{21}r_{21}=-i\left[G_{21}(r_{1}-r_{2})-
G_{01}r_{20}\right],&\nonumber\\
&P_{02}r_{02}=-i\left[G_{01}r_{12}-G_{12}r_{01}\right];&\\
&\Gamma_{1}r_{1}=2\Re[iG_{12}r_{21}+iG_{10}r_{01}],\,
\Gamma_{2}r_{2}=-2\Re[iG_{12}r_{21}]+\gamma_{2}r_{1},&\nonumber\\
&  r_{0}+r_{1}+r_{2}=1;&\label{eq1}\\
&P_{23}r_{23}=-i\left[-G_{23}r_{2}+
G_{21}r_{13}-G_{03}r_{20}\right],&\nonumber\\
&P_{03}r_{03}=-i\left[-G_{03}r_{0}+
G_{01}r_{13}-G_{23}r_{02}\right],\nonumber& \\
&P_{13}r_{13}=-i\left[G_{12}r_{23}-G_{23}r_{12}+
G_{10}r_{03}-G_{03}r_{10}\right]\, &
\end{eqnarray}
Here $r_{ij}$ and $r_i$ are the amplitudes of the off-diagonal and diagonal elements
of the density matrix, $\displaystyle P_{ij}=\Gamma_{ij}+ i(\Omega_{l}-k_{l}v)$,
$\displaystyle \Omega_{l}$  is the resonance detuning for the corresponding resonant
field (e.g., $\displaystyle \Omega_1=\omega_1-\omega_{01}$), $\displaystyle
\Gamma_{ij}$ are homogeneous half-widths of the transitions (coherence relaxation
rates), $v$ is the projection of the atom velocity on the direction of the wave
vectors ${\bf k}_l$, $\displaystyle \gamma_{1,2}$ are relaxation rates from level $\it
1$ to $\it 0$ and from $\it 1$ to $\it 2$, accordingly,  $\displaystyle
\Gamma_{1},\Gamma_{2}$ are relaxation rates of the populations of the levels $\it 1$
and $\it 2$ accordingly, $\displaystyle G_{ij}=-{E_j d_{ij}}/{2\hbar}$ are the
 Rabi frequencies, and $d_{ij}$ are the electric dipole momenta of the
transitions. Only the resonant couplings are accounted for.

The solution for the coherence $r_{02}$ and for those at the
allowed transitions, which determine  absorption and refraction at
the frequencies $\omega_1$, $\omega_2$, $\omega_3$, are found in
the form
\begin{eqnarray*}
& &r_{02}=\frac{G_{01}G_{12}}{\tilde P_{02}}\left\{ \frac{\Delta
r_{10}}{P_{01}}-\frac{\Delta r_{21}}
{P_{21}^{*}}\right\}=\frac{G_{01}G_{12}}{P_{01}P_{21}^{*}}R_{02},\\
& &r_{01}=-i\frac{G_{01}}{P_{01}}\left\{\Delta
r_{10}-\frac{|G_{12}|^2} {\tilde P_{02}}\left(\frac{\Delta
r_{10}}{P_{01}}-\frac{\Delta r_{21}}
{P_{21}^{*}}\right)\right\}=-i\frac{G_{01}}{P_{01}}R_1,\\
& &r_{12}=-i\frac{G_{12}}{P_{21}^{*}}\left\{\Delta
r_{21}+\frac{|G_{01}|^2} {\tilde P_{02}}\left(\frac{\Delta
r_{10}}{P_{01}}-\frac{\Delta r_{21}}
{P_{21}^{*}}\right)\right\}=-i\frac{G_{12}}{P_{21}^{*}}R_2,\\
&
&r_{23}=i\frac{G_{23}}{P_{23}}\left\{r_{2}\left(1-\frac{|G_{12}|^2}{P_{23}
\tilde P_{13}}\right)-\frac{|G_{01}G_{12}|^2}{P_{03}\tilde P_{13}}
\frac{R_{02}}{P_{01}P_{21}^{*}}
-\frac{|G_{12}|^2}{P_{21}^{*}\tilde P_{13}}R_{2}\right\},
\end{eqnarray*}
where $\Delta r_{ij}$ are intensity-dependent population differences, and $\tilde
P_{02}$ and $\tilde P_{13}$ are two-photon denominators ``dressed'' by the driving
fields:
\begin{eqnarray}\label{Den}
\tilde P_{02}=P_{02}+\frac{|G_{01}|^2}{P_{21}^{*}}+\frac{|G_{12}|^2}{P_{01}},\quad
\tilde P_{13}=P_{13}+\frac{|G_{01}|^2}{P_{03}}+\frac{|G_{12}|^2}{P_{23}}.
\end{eqnarray}

The coherence $r_{03}$ consists of two terms. One, $\bar r_{03}$, determines
absorption and refraction at the frequency $\omega_S$, and the other, $\tilde
r_{03}$, determines FWM at $\omega_1-\omega_2+\omega_3=\omega_S$:
\begin{eqnarray}\label{Chi}
& &\bar
r_{03}=i\frac{G_{03}}{P_{03}}\left\{r_{0}\left(1-\frac{|G_{01}|^2}{P_{03}
\tilde P_{13}}\right)-\frac{|G_{01}G_{12}|^2}{P_{23}\tilde P_{13}}
\frac{R_{02}^{*}}{P_{01}^{*}P_{21}}
+\frac{|G_{01}|^2}{P_{01}^{*}\tilde P_{13}}R_{1}^{*}\right\},\label{Chi1}\\
& &\tilde
r_{03}=-i\frac{G_{01}G_{12}G_{23}}{P_{01}P_{21}^{*}P_{03}}
\left\{R_{02}\left(\frac{|G_{01}|^2}{P_{03}\tilde P_{13}}
-1\right)+\frac{P_{01}P_{21}^{*}}{\tilde P_{13}}
\left(\frac{r_2}{P_{23}}+\frac{R_2}{P_{21}^{*}}\right)\right\},\label{Chi2}
\end{eqnarray}
where $\displaystyle P_{03}=\Gamma_{03}+i(\Omega_{1}-\Omega_{2}+\Omega_{3}- k_{S}v)$
and $\quad k_{S} = k_1 - k_2 + k_3.$

Making use of the above equations, the solution for the populations
can be found from (\ref{eq1}) as
\begin{eqnarray}
r_{2}=\frac{Y_{1}W_{02}-Y_{3}W_{01}}{Y_{1}Y_{4}-Y_{3}Y_{2}}, \quad
r_{1}=\frac{Y_{4}W_{01}-Y_{2}W_{02}}{Y_{1}Y_{4}-Y_{3}Y_{2}}, \quad
r_{0}=1-r_{1}-r_{2},\label{r}
\end{eqnarray}
where
\begin{eqnarray*}
&Y_{1}=2W_{01}-W_{02}+\gamma_{1},\quad Y_{2}=W_{02}+W_{01}+\Gamma_{2},&\\
&Y_{3}=W_{02}-W_{21}-\gamma_{2},\quad Y_{4}=W_{02}+W_{21}+\Gamma_{2},&
\end{eqnarray*}
and
\begin{eqnarray*}
&W_{01}=2|G_{01}|^{2}\Re\left\{\left(1-\displaystyle\frac{|G_{12}|^2}{P_{01}\tilde
P_{02}} \right)/P_{01}\right\},&\\
&W_{21}=2|G_{12}|^{2}\Re\left\{\left(1-\displaystyle\frac{|G_{01}|^2}{P_{21}^{*}\tilde
P_{02}} \right)/P_{21}^{*}\right\},&\\
&W_{02}=2|G_{01}|^{2}|G_{12}|^{2}\Re\left\{{1}/{P_{01}P_{21}^{*}\tilde
P_{02}}\right\}.&
\end{eqnarray*}
\subsection{Coherence-induced Doppler-free resonance}

The appearance of a DF resonance and therefore the coupling of molecules from a wide
velocity interval can be understood as follows. \textcolor{red}{Ac-Stark shifts}
$\Im\left\{{|G_{01}|^2}/{P_{21}^{*}}\right\}$ and $\Im\left\{
{|G_{12}|^2}/{P_{01}}\right\}$ in (\ref{Den}), \textcolor{red}{originating from the
coherence $r_{02}$, depend on the radiation intensity and frequency detunings. The
later ones, in turn, depend on Doppler shifts.  This allows one, by making judicious
choice of the intensities and detunings of the driving fields, to ``draw" into the
dressed two-photon resonance {\it all} the molecules, independent of their
velocities.} In the limiting case, when the detuning from the intermediate resonance
is much greater than the Doppler HWHM of the allowed optical transitions, the
modified two-photon resonance described by the denominator $\tilde P_{02}$ can be
presented in the lowest order in $k_{1,2}v/\Omega_{1,2}$ as
\begin{equation}\label{P02}
\tilde P_{02}=\tilde\Gamma_{02}+i\tilde\Omega_{02}-i
\left\{\left(1+\frac{|G_{12}|^2}{\Omega_1^2}\right)k_1-
\left(1+\frac{|G_{01}|^2}{\Omega_2^2}\right)k_2\right\}v,
\end{equation}
where $\displaystyle{\tilde\Gamma_{02}}=\Gamma_{02}+\frac{|G_{01}|^2}{\Omega_2^2}
\Gamma_{12}+\frac{|G_{12}|^2}{\Omega_1^2}\Gamma_{01},\quad \tilde\Omega_{02}=\Omega_1
- \Omega_2 +\frac{|G_{01}|^2}{\Omega_2}- \frac{|G_{12}|^2}{\Omega_1}$ are the power
broadened HWHM and ac-Stark shifted two-photon resonance detuning. From (\ref{P02})
it follows that the requirements for the induced Doppler-free resonance to be
achieved are $\displaystyle
\tilde\Omega_{02}=0$\  and \ $\displaystyle\tilde{k_1}=\tilde{k_2}$, where\\
$\displaystyle\tilde{k_1}v=\left(1+\frac{|G_{12}|^2}{\Omega_1^2}\right)k_1v$ and $
\displaystyle\tilde{k_2}v=\left(1+\frac{|G_{01}|^2}{\Omega_2^2}\right)k_2v$\\
are intensity-dependent Doppler shifts. Eventually the equations take form
\begin{eqnarray}
&k_1\left(1+\displaystyle\frac{|G_{12}|^2}{\Omega_1^2}\right)-
k_2\left(1+\displaystyle\frac{|G_{01}|^2}{\Omega_2^2}\right)=0,&\label{eq3}\\
&\Omega_1\left(1-\displaystyle\frac{|G_{12}|^2}{\Omega_1^2}\right)-
\Omega_2\left(1-\displaystyle\frac{|G_{01}|^2}{\Omega_2^2}\right)=0.&\label{eq4}
\end{eqnarray}
From (\ref{eq3})-(\ref{eq4}) follows a cubic equation as $\Omega_2$:
\begin{equation}\label{kub}
\Omega_{2r}^3 - (2-K) \Omega_{2r}^2\Omega_1- \Omega_{2r}|G_{01}|^2
+ K\Omega_1|G_{01}|^2=0,\quad K=k_2/k_1,
\end{equation}
which determines the detuning of  field $E_2$, corresponding to
the induced resonance, under given values of the Rabi frequency and
detuning of field $E_1$. Here, the Rabi frequency of $E_2$
must be
\begin{equation}\label{G2}
|G_{12}|^2 = \Omega_1^2\left\{K\left(1+\frac{|G_{01}|^2}
{\Omega_{2r}^2}\right)-1\right\}.
\end{equation}
One of the roots of the equation (\ref{kub}) corresponds to the
dressed DF two-photon resonance,
\begin{equation}
\Omega_{2r}=(A^{1/3}+9B+2\Omega_{1}-\Omega_{1}K)/3,
\end{equation}
where
$\displaystyle A=9\Omega_{1}|G_{01}|^{2}(1-2K)+
\Omega_1^3(8-12K+6K^2-K^3)+3[3\Omega_1^2|G_{01}|^4\times$
\begin{center}
$\displaystyle \times(-1-8K+11K^2)+3\Omega_1^4|G_{01}|^2
(-8K+12K^2-6K^3+K^4)-3|G_{01}|^6]^{1/2}$,\\
\end{center}
\begin{center}
$\displaystyle
B=(|G_{01}|^2+4\Omega_1^2-4\Omega_1^2K+\Omega_1^2K^2)/3A^{1/3},\quad
K={k_2}/{k_1}.$
\end{center}
The conditions for the CPT and DF resonances may fit each other or differ, depending
on the specific case. This and corresponding outcomes regarding FWM will be
investigated below numerically for the most optimum situations, where the analytical
solution can not be obtained.
\section{Numerical analysis}
The graphs presented below are computed numerically based on averages over the
velocity equations (\ref{Chi1}), (\ref{Chi2}) and (\ref{r}). For the numerical
analysis we have used the parameters of the sodium dimer transition with the following
wavelengths \cite{BAB}: $X^1\Sigma_g^+(v''=3,J''=43)$ -- $B^1\Pi_u(6,43)
(\lambda_{01}= 488$ nm) -- $X^1\Sigma_g^+(13,43)$ $(\lambda_{21} = 525$  nm) --
$A^1\Sigma^+_u(24,44)(\lambda_{23}= 655$  nm) -- $X^1\Sigma_g^+(3,43) (\lambda_{13}=
598$ nm). The corresponding homogeneous half-widths of the transition are 20.69,
23.08, 18.30 and 15.92 MHz, whereas the Doppler half-widths $\Delta\omega_{iD}$ are
equal to 0.92, 0.85, 0.68 and 0.75 GHz. The numerical simulations allow us to analyze
the velocity-averaged equation for the detunings, where the approximation taken in
(\ref{P02}) is not valid.
\subsection{Nonlinear resonances in FWM polarizations, in absorption indices
 and in the level populations}
\begin{figure}[!h]
\begin{center}
\includegraphics[width=.45\textwidth]{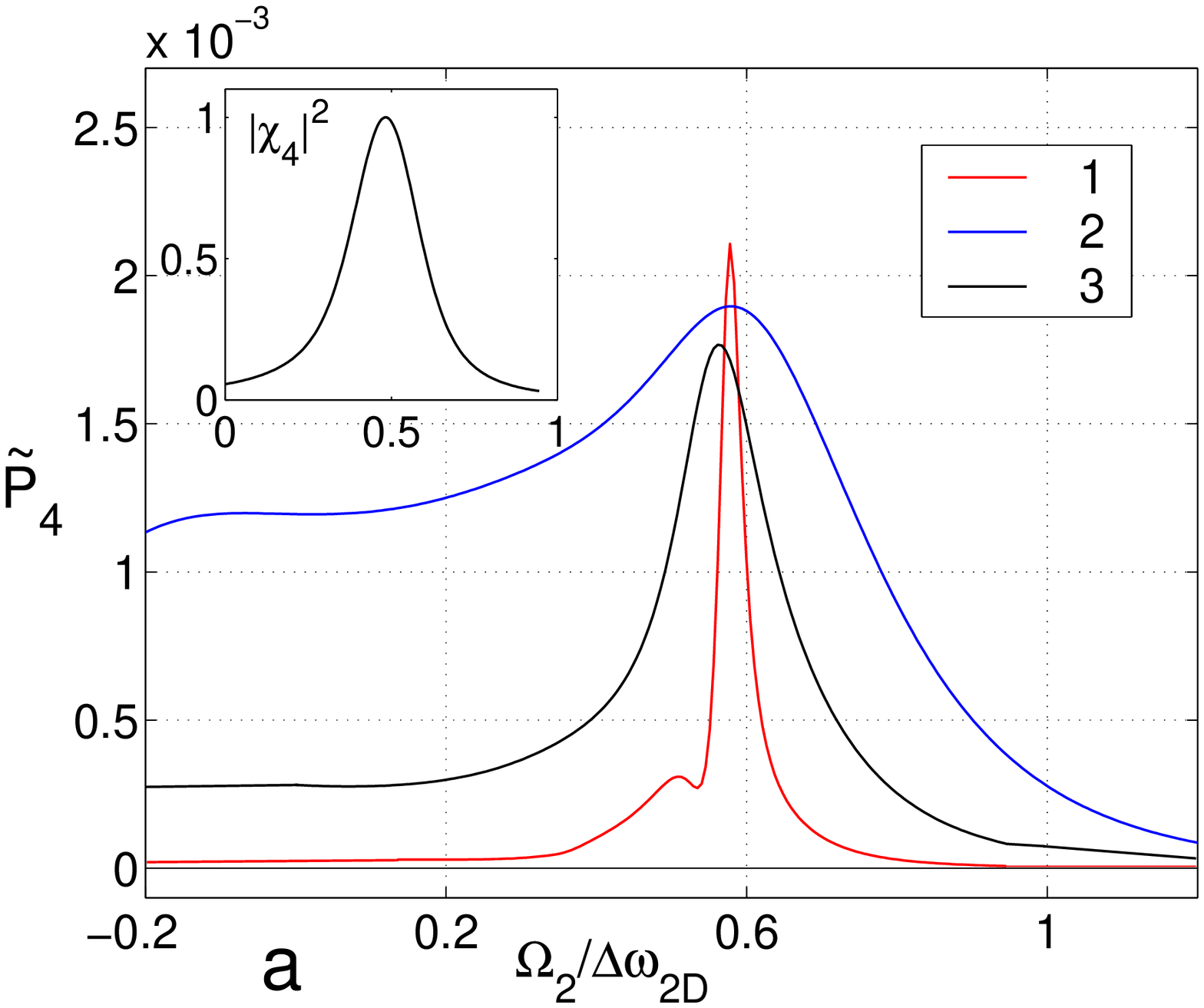}
\includegraphics[width=.45\textwidth]{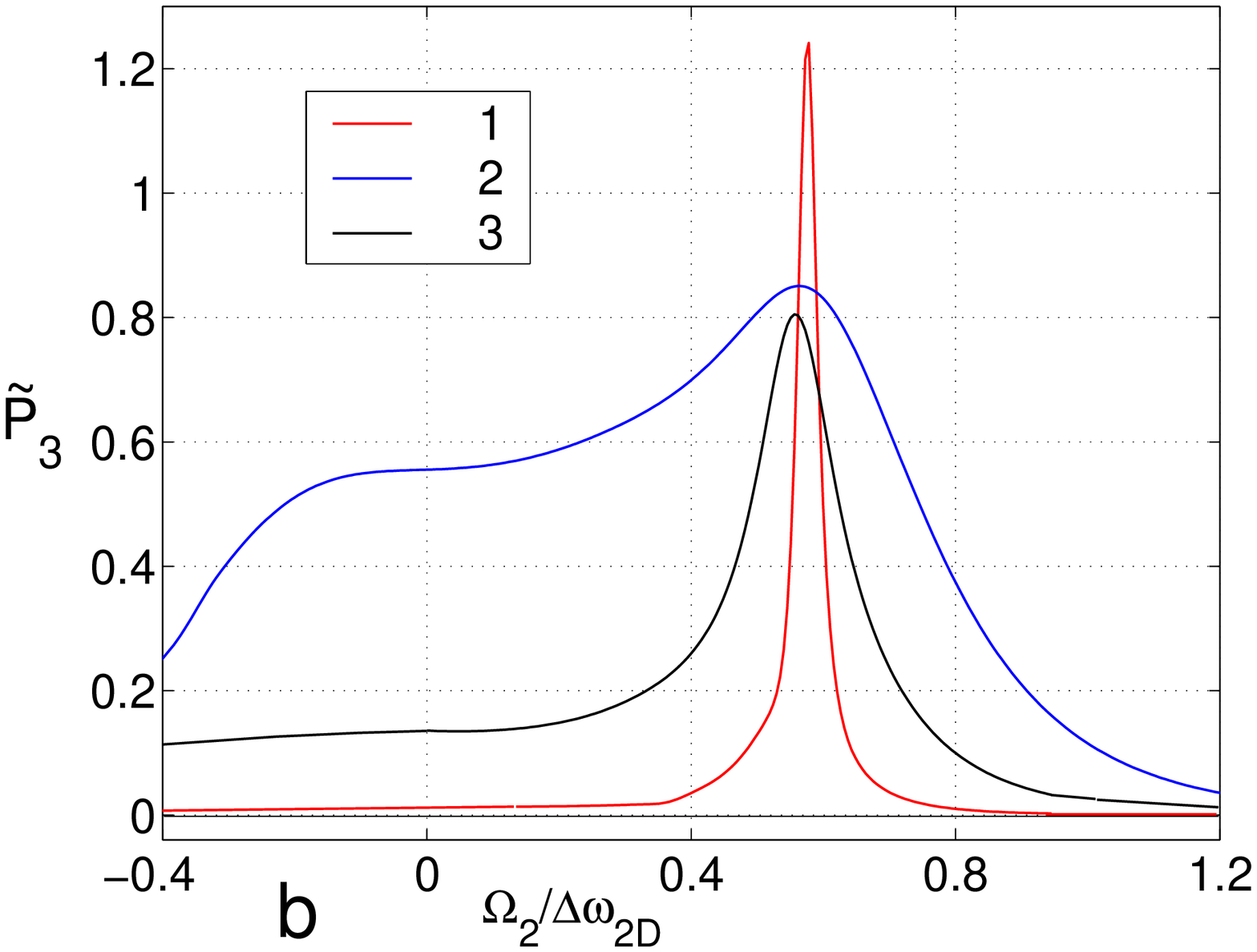}
\end{center}
\vspace{-10pt}\caption{\label{figPom} Normalized squared module of the FWM
polarizations $\tilde P_4$ ({a}) and $\tilde P_3$ ({b}) vs the detuning of driving
field $E_2$. The plots correspond to: 1-- DF resonance, 2 -- CPT, 3 -- intermediate
case, subplot -- same resonance, but under weak fields.}\vspace{-10pt}
\end{figure}
Figure ~\ref{figPom}, plots 1 \textcolor{red}{illustrate the coherence-induced
compensation of Doppler shifts with ac-Stark shifts resulting in the resonance
narrowing in the squared module of the velocity-averaged reduced nonlinear FWM
polarizations}

\noindent $|\tilde P_4|^2 = (|\tilde\chi_4|^2/|\tilde\chi_{40}|^2) |g_{01}|^2
|g_{12}|^2$ [Fig.\ref{figPom} ({\bf a})] and $\displaystyle |\tilde
P_3|^2=(|\tilde\chi_3|^2/|\tilde\chi_{40}|^2) |g_{01}|^2 |g_{12}|^2$ [Fig.\ref{figPom}
({\bf b})]. The nonlinear susceptibility is reduced by its maximum value in the same
frequency range, but for negligibly-weak $E_1$ and $E_2$ fields,
$\displaystyle{g_{01}}=G_{01}/\Delta\omega_{1D}$,
$\displaystyle{g_{12}}=G_{12}/\Delta\omega_{2D}$. \textcolor{red}{A substantial
narrowing is seen from comparison with the subplot}, presenting the spectral
dependence of the velocity-averaged squared module of the same susceptibility, but in
the negligibly weak fields, normalized to unity.  The HWHM of the resonance in the
subplot is approximately 70 MHz, which corresponds to the Doppler width of the
nonperturbated Raman transition, whereas the HWHM of the resonance in plots 1 is 15
MHz. This indicates some power and residual Doppler broadening. The Rabi frequencies
of the fields $E_1$ and $E_2$ are equal to 157 MHz and 85 MHz. The detuning of $E_1$
is equal to 413.8 MHz ($0.45\cdot\Delta\omega_{1D}$), and detuning of $E_3$ is equal
to  -695.5 MHz ($-1.02\cdot\Delta\omega_{3D}$). Plots 2 display the same dependences,
but related to the CPT conditions (the half-width of this resonance in $|\tilde
P_4|^2$ is 200 MHz). In this case, the required Rabi frequencies of $E_1$ and $E_2$
are equal to 351 MHz and 332 MHz, respectively, while the detunings are identical to
those in plots 1. Plots 3 display the same dependence, but under the intensities,
that provide narrowing in the intermediate range (HWHM of resonance is 70 MHz). In
this case, the Rabi frequencies of the fields $E_1$ and $E_2$ are equal to 222 MHz
and 235 MHz, respectively, whereas the detunings are the same as in the previous
plots. \textcolor{red}{The important outcome is that in the Doppler-free regime,
larger FWM polarizations can be accomplished under lower driving fields.} Also, we
want to stress the substantial difference in magnitudes of the nonlinear
susceptibilities  $\chi_4$ and $\chi_3$, which follows from the interference of
contributions of molecules in different velocity intervals to the macroscopic
nonlinear polarizations.
\begin{figure}[!h]
\begin{center}
\includegraphics[width=.45\textwidth]{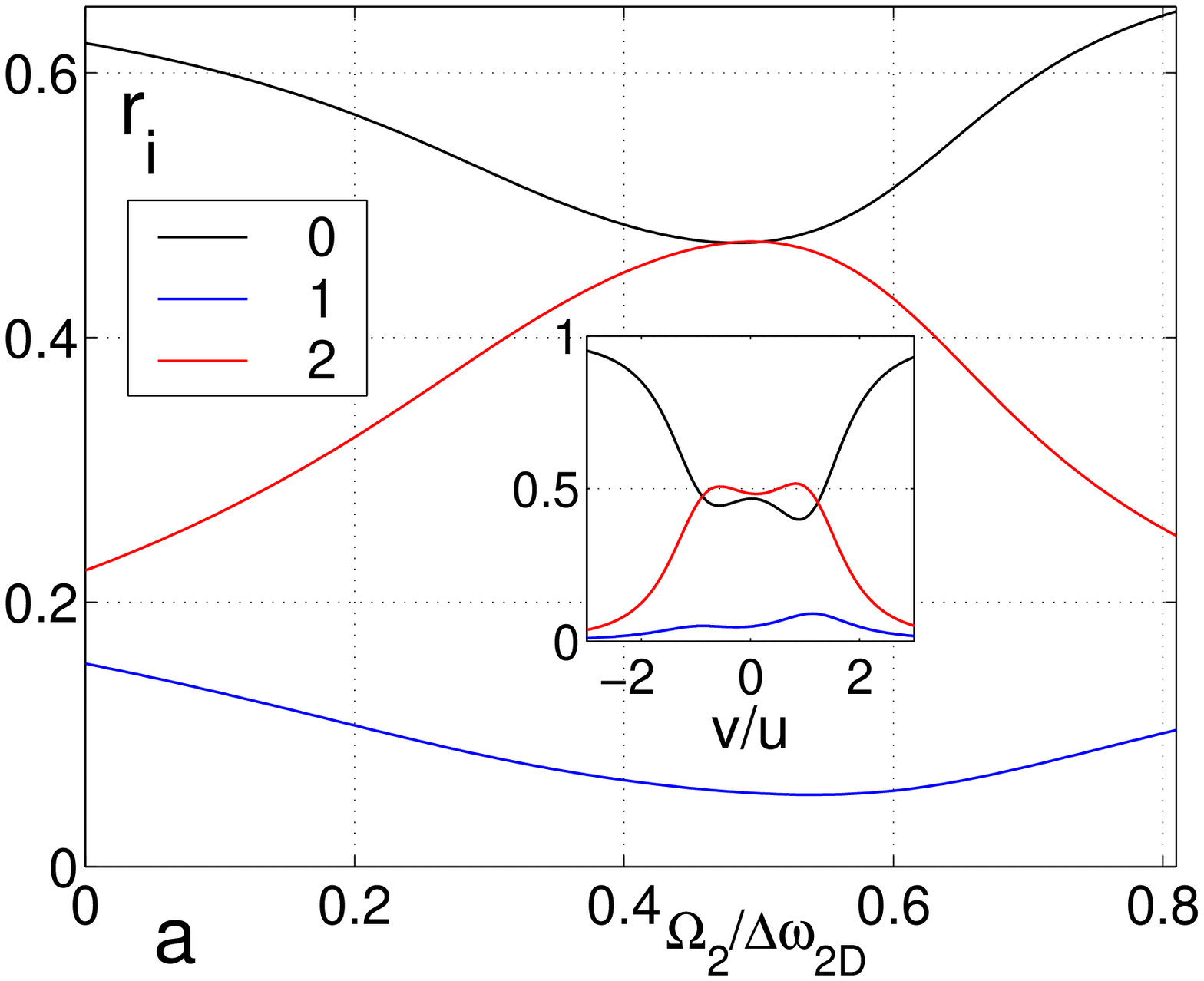}
\includegraphics[width=.45\textwidth]{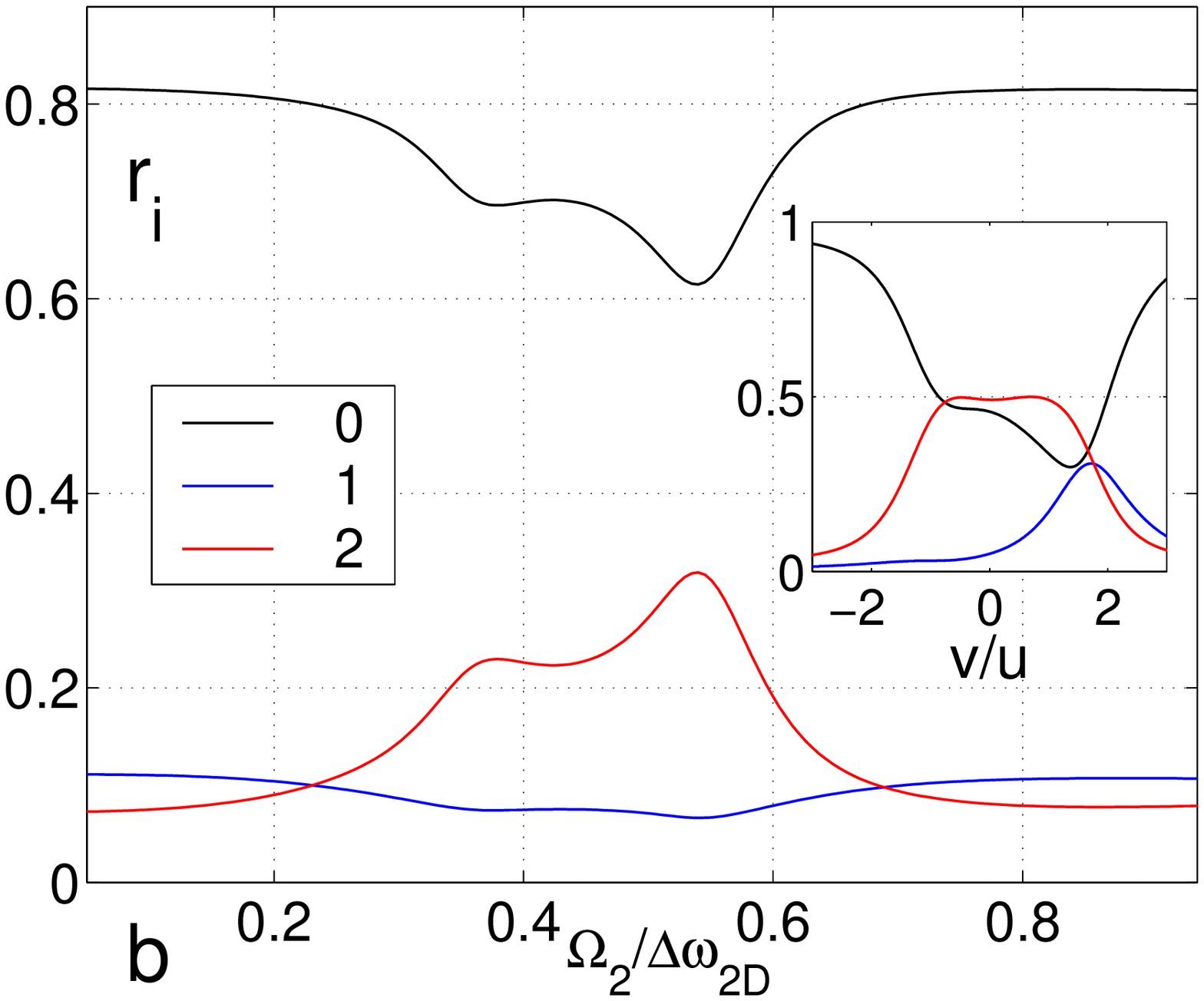}
\end{center}
\vspace{-10pt}\caption{\label{figPOP}Velocity integrated population of the energy
levels  vs detuning of driving field $E_2$. {a}-- near CPT and CPT regime, subplot --
velocity distribution of the coherently trapped populations; {b}-- near DF regime,
subplot -- velocity distribution of populations under DF regime.}\vspace{-10pt}
\end{figure}
Figure \ref{figPOP}(a) shows dependence of the velocity integrated populations on the
detuning  of the field $E_2$, while the Rabi frequencies  of $E_1$ and $E_2$ are
equal to 351 MHz and 332 MHz, respectively so that the CPT conditions can be
fulfilled under appropriate detuning. The subplot  displays the distribution of the
populations over velocities, whereas the detuning corresponds to the CPT regime for
the velocity-integrated populations. Here $u$ is the thermal velocity with Maxwell
envelope removed. \textcolor{red}{The subplot indicates CPT for molecules in $v=0$;
however, even inversion of the populations occurs in a relatively wide velocity
intervals.}

Figure \ref{figPOP}(b) is computed for the Rabi frequencies
of the fields $E_1$ and $E_2$ equal to 157 MHz and 85 MHz, respectively.
The inset displays the distribution of the populations over velocities
at the detuning corresponding to DF. The distribution is even more
complicated compared to that in the previous case.

\begin{figure}[!h]
\begin{center}
\includegraphics[width=.45\textwidth]{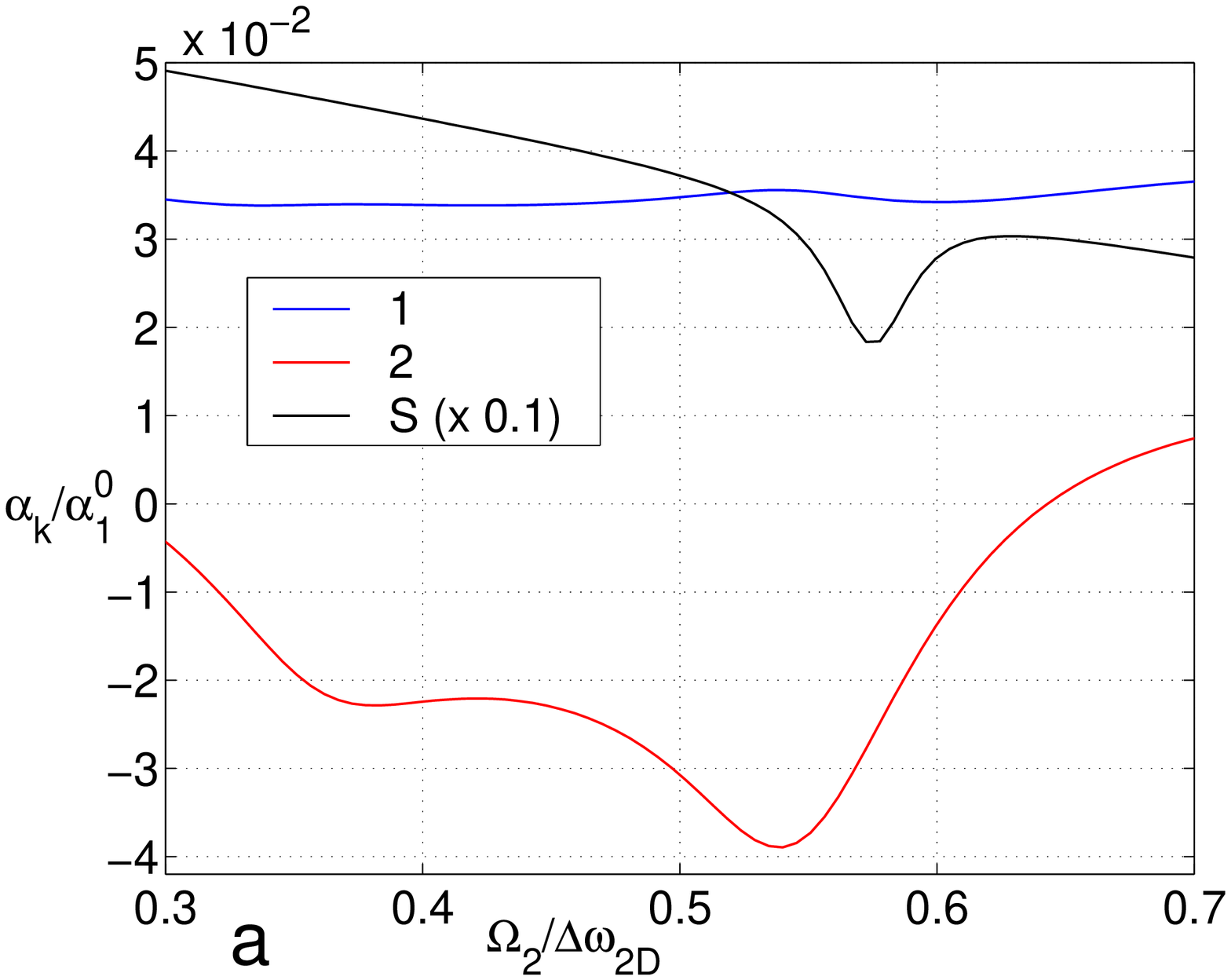}
\includegraphics[width=.45\textwidth]{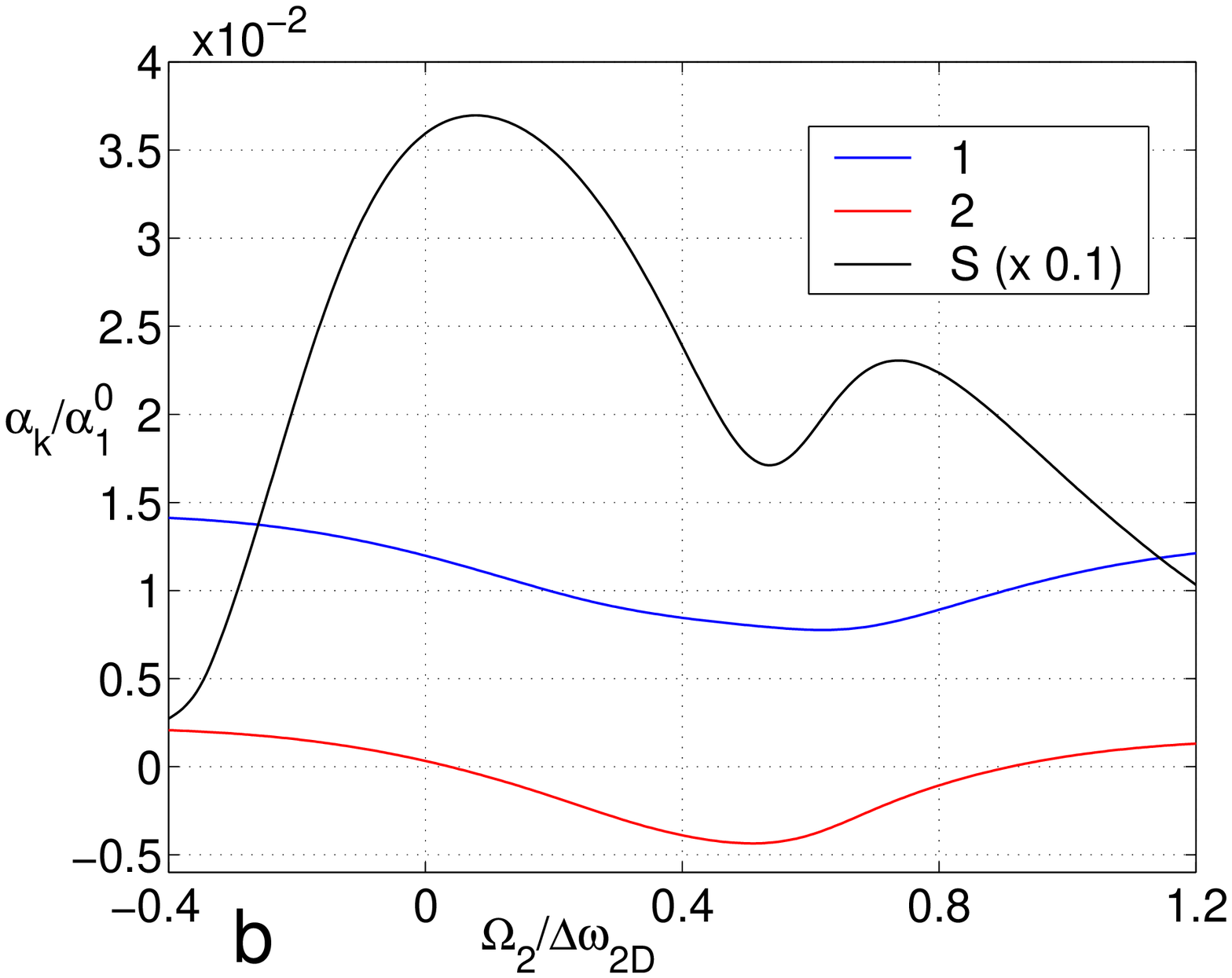}
\end{center}
\vspace{-10pt}\caption{\label{figAlp}Velocity averaged absorption indices
$\alpha_{1,2,S}$ near DF (a) and CPT (b) regimes vs detuning of the driving field
$E_2$. }
\end{figure}

Figure \ref{figAlp} displays velocity-averaged absorption indices
vs detuning of the field $E_2$.
The intensities of the fields are the same as in the corresponding graphs given
above. The graph~\ref{figAlp} (a) shows a substantially decreased absorption index for
the first field due to saturation effects, which is almost independent of the
detuning of the second field because of its relative weakness. The second field
experiences Stokes gain. The plot for the generated beam (reduced by 10) displays a
nonlinear resonance in the absorption index. With growth of the second field
intensity, so that CPT becomes possible, both absorption of the first and gain of the
second driving field experience substantial decrease in the CPT regime. Absorption of
the generated field dramatically changes as well (Figure~\ref{figAlp} (a)).

A linear approximation in $k_{1,2}v/\Omega_{1,2}$ (see equation (\ref{P02})) is valid
for large detunings from the intermediate resonance. While for tuning close to the
resonance, the concurrently coupled velocity interval decreases, the nonlinear
response of the molecules grows. At the same time, absorption of the coupled
radiations increases as well.

The output FWM generated radiation is determined by the interplay
of the above considered processes. Moreover, resonant FWM may not
be viewed as a sequence of the independent elementary acts of
absorption, gain and FWM. Interference of these elementary
processes plays a crucial role, even for a qualitative prediction of
the spatial dynamics of the generated field \cite{TAR}. Therefore,
appropriate optimization is required in order to achieve the
maximum FWM-generation output. In the next section, relevant
results are presented based on the above analyzed dependencies.
\subsection{Coherent quantum control of FWM in a double-$\Lambda$
Doppler broadened medium}
The solution of the Maxwell equations for the coupled traveling
waves can be found in the form
\begin{equation}
E^{j}(z,t)=\Re\{E_{j}(z)\exp [{\rm i}(\omega _{j}t-k_{j}z)]\},
\end{equation}
where $k_{j}$ are complex wave number at the corresponding
frequencies, $k_j = k_j'-{\rm i}\alpha _{j}/2$. The set of the
equations for the amplitudes, relevant to the case under
consideration, is
\begin{eqnarray}
\frac{d\,E_1}{dz}=-\frac{\alpha_1}{2} E_1, \quad
\frac{d\,E_2}{dz}=-\frac{\alpha_2}{2} E_2,\label{eq_m1}\\
\frac{d\,E_3}{dz}=-\frac{\alpha_3}{2}E_3 +
\sigma_3 E_1^*E_2 E_S \exp\{-i\Delta kz\},\\
\frac{d\,E_S}{dz}=-\frac{\alpha_S}{2}E_S+ \sigma_S E_1 E_2^* E_3
\exp\{i\Delta kz\},\label{eq_m2}
\end{eqnarray}
\noindent where $\sigma_3=i 2 \pi k_3 \tilde{\chi_3}$ and $\sigma_S=i
2 \pi k_S \tilde{\chi_S}$ are cross coupling parameters,
$\tilde{\chi_3}$ and $\tilde{\chi_S}$ are intensity-dependent
nonlinear susceptibilities for the FWM processes $\omega
_{S}\leftrightarrow \omega _{1}-\omega _{2}+\omega _{3}$ and $\Delta
k=k_{S}-k_{1}+k_{2}-k_{3}$. The boundary conditions are
$E_{l}(z=0)=E_{0l}$ ($l=1,2,3$) and $E_{S}(z=0)=0$. We assume that
intensities of the weak waves with frequencies $\omega_3$ and
$\omega_S$ are weak enough so that the change of the strong
fields $E_{1,2}$ due to FWM conversion can be
neglected. The quantum efficiency of conversion QEC of the
radiation  $E_3$ in $E_{S}$ is defined by the equation
\begin{equation}\label{qu1}
\eta_{{\rm
q}}=(\omega_3/\omega_S)|E_S(z)/E_3(0)|^2\exp(-\alpha_Sz).
\end{equation}
Assuming that  the condition of phase matching $\Delta k'=0$ can
be ensured (e.g., with a buffer gas), the solution for  QEC of
the above given equations can be expressed as
\begin{eqnarray}\hspace*{-5mm}
&\eta_{q}=\displaystyle\frac{\omega_{3}}{\omega_{S}}
\bigl|\displaystyle\frac{\sigma_{S}}{ g_{0}}\bigl|^{2}
\bigl|E_{01}E_{02}\bigl|^2\exp\left[(\Re g_{0}-\frac{\alpha}{2})z\right]
\bigl|1-\exp(-g_{0}z)\bigr|^{2},&\nonumber \\
&g_{0}=\left[\left[
(\alpha_{1}+\alpha_{2}+\alpha_{3}-\alpha_{S}/2)\right]^{2}+
4\sigma_{S}\sigma_{3}\bigl|E_{01}E_{02}\bigl|^2\right]^{1/2},
\quad \alpha=\alpha_{3}+\alpha_{S}.& \nonumber
\end{eqnarray}
These formulas account for nonlinear resonances both in FWM nonlinear polarizations
and in absorption indices. In order to compute the absolute magnitude of QEC (under
the assumption of $\Delta k'=0$) we have used the data for the Frank-Condon factors of
0.068, 0.142, 0.02 and 0.036 for the transition with wavelengths $\lambda_{01}$,
$\lambda_{21}$, $\lambda_{23}$ and $\lambda_{03}$ accordingly.

Figure \ref{figQEC} presents a numerical simulation of the evolution of QEC along the
vapor cell. The distance is scaled to the resonance optical density $\alpha_{01}z$
with the driving fields being turned off. The plot 1 corresponds to the case where
the conditions for Doppler compensation at the transition 0--2 are fulfilled on the
entrance of the cell. The Doppler-free resonance would be narrower if one of the
coupled fields were weak. However, that would give rise to larger absorption and to
decreased nonlinear FWM polarization. The optimization accounting for the change of
the driving fields along the medium also shows that it does not substantially change
the optimum input intensity values for the plots presented in Fig. \ref{figQEC}. Plot
2 displays the same dependence, but for the optimum CPT conditions. Alternatively,
they are not optimum for the elimination of Doppler broadening.
\begin{floatingfigure}{0.5\textwidth}
\begin{center}
\includegraphics[width=.45\textwidth]{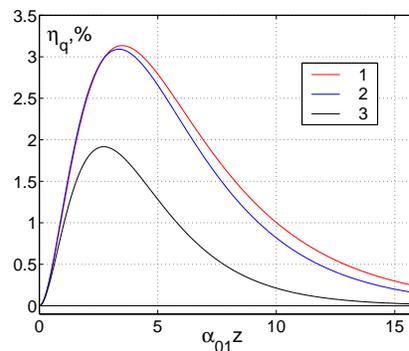}
\end{center}
\vspace{-25pt}\textcolor{black}{\caption{\label{figQEC}{\em QEC} vs optical density of
the medium. 1 -- DF, 2 -- CPT, 3 -- intermediate regimes.}}\vspace{-10pt}
\end{floatingfigure}
\noindent\textcolor{red}{The maxima of these two curves are comparable, but in the
Doppler- free regime, quite lower intensities of the fundamental radiations are
required.} Plot 3 displays a similar dependence in the intermediate case. All
intensities and detunings are identical to those used for computing the previous
figures.

For the transitions under consideration, the Rabi frequency on the level of 100 MHz
corresponds to powers of about 100 mW focused on a spot on the order of $10^{-4}$
cm$^2$, which can be realized with common cw lasers with the confocal parameter of
focusing of about 2.5 cm. At a temperature of about 700 K, the optimum optical
density of the dimers is achievable for vapor lengths of about 2 cm. This is in
accord with the parameters of typical cw FWM experiments (see, e.g. \cite{BAB}).

The above indicated intensity corresponds to about $10^{11}$ photons per cubic
centimeter. Accounting for the molecule number density at the indicated temperature,
which is on the order on $10^{13}$ cm$^{-3}$, \textcolor{red}{we have about $10^{-2}$
photon/molecule.}
\section{Conclusion}
Coherent control of populations \cite{BER} and of four-wave mixing \cite{HER} with
pulses shorter than the dephasing time $T_2$ has proven to be a powerful tool for
manipulating nonlinear-optical and chemical properties of free atoms and molecules.
In these cases, maximum coherence can be achieved as a result of Rabi oscillations of
the two-photon transition, and the required driving intensity is much higher than that
proposed in this paper. Consequently, Doppler broadening of the coupled transitions
does not play an important role.

On the contrary, this paper considers coherent quantum control of resonant four-wave
mixing processes, where Doppler-broadening of a double-$\Lambda$ medium is the factor
of crucial importance. An approach enabling one to enhance the efficiency of nonlinear
optical conversion and, at the same time, to decrease the required intensity for the
fundamental pump beams is proposed. This approach is based on the elimination of
Doppler broadening of the resonant two-photon transition. The advantages of the
proposed method as compared to those based on coherent population trapping, where
inhomogeneous broadening may play important role too, are illustrated with numerical
simulations for realistic experimental schemes. The results obtained contribute to
the field of nonlinear optics at the level of several photons per atom, which is
currently attracting growing interest.
\section{Acknowledgments}
TFG and AKP  thank U.S. National Research Council - National Academy of Sciences for
support of this research through the international Collaboration in Basic Science and
Engineering (COBASE) program. AKP and ASB acknowledge funding support from the
International Association of the European Community for the promotion of co-operation
with scientists from the New Independent States of the former Soviet Union (INTAS)
(grant INTAS-99-19) and Russian Foundations for Basic Research  (grant 99-02-39003)
and from the Center on Fundamental Natural Sciences at St. Petersburg University
(grant 97-5.2-61).

\end{document}